\documentclass[%
 twocolumn,
 superscriptaddress,
 amsmath,amssymb,
 aps,
 prl,
]{revtex4-2}

\usepackage{float}
\usepackage{subfigure} 
\usepackage{bm}                 
\usepackage{graphicx}
\usepackage{dcolumn}
\usepackage{bm}
\usepackage[colorlinks,citecolor=blue,linkcolor=blue,urlcolor=blue]{hyperref}

\begin{document}

\title{Coherent XUV supercontinuum emission from atomic bound states} 
\author{Jing Zhao}
\thanks{These authors contributed equally to this work.}
\affiliation{Department of Physics, National University of Defense Technology, Changsha 410073, China}
\author{Xiaowei Wang}%
\thanks{These authors contributed equally to this work.}
\affiliation{Department of Physics, National University of Defense Technology, Changsha 410073, China}
\author{Li Wang}%
\thanks{These authors contributed equally to this work.}
\affiliation{Department of Physics, National University of Defense Technology, Changsha 410073, China}%
\author{Jiacan Wang}%
\affiliation{Department of Physics, National University of Defense Technology, Changsha 410073, China}
\author{Yalei Zhu}%
\affiliation{Department of Physics, National University of Defense Technology, Changsha 410073, China}
\author{Fan Xiao}%
\affiliation{Department of Physics, National University of Defense Technology, Changsha 410073, China}
\author{Wenkai Tao}%
\affiliation{Department of Physics, National University of Defense Technology, Changsha 410073, China}
\author{Zhigang Zheng}%
\affiliation{Department of Physics, National University of Defense Technology, Changsha 410073, China}
\author{Haizhong Wu}%
\affiliation{Department of Physics, National University of Defense Technology, Changsha 410073, China}
\author{Xu Sun}%
\affiliation{Department of Physics, National University of Defense Technology, Changsha 410073, China}
\author{Yue Lang}%
\affiliation{Department of Physics, National University of Defense Technology, Changsha 410073, China}
\author{Congsen Meng}%
\affiliation{Department of Physics, National University of Defense Technology, Changsha 410073, China}
\author{Dongwen Zhang}%
\affiliation{Department of Physics, National University of Defense Technology, Changsha 410073, China}
\author{Zhihui Lv}%
\affiliation{Department of Physics, National University of Defense Technology, Changsha 410073, China}
\author{Jinlei Liu}%
\email{liujinlei@nudt.edu.cn}
\affiliation{Department of Physics, National University of Defense Technology, Changsha 410073, China}
\author{Zengxiu Zhao}
\email{zhaozengxiu@nudt.edu.cn}
\affiliation{Department of Physics, National University of Defense Technology, Changsha 410073, China}
\affiliation{Hunan Key Laboratory of Extreme Matter and Applications, National University of Defense Technology, Changsha 410073, China}

\date{\today}

\begin{abstract}
Coherent supercontinuum radiation in the extreme-ultraviolet (XUV) range is indispensable for synthesizing attosecond light pulses and for exploring transient atomic structures. Here, we report the striking observations of coherent XUV supercontinuum (XSC) extended from below to far above the ionization threshold, which exhibits completely different temporal and spatial properties comparing to the conventional rescattering induced high harmonic generation (HHG).  We demonstrate that the strong-field created coherence among bound orbitals strongly distort the atomic transition energies during the pulse, leading to coherent emission spanning tens of electron-volts, in contrast to the line emission via free-induction decay occurring after the pulse. The supposed non-radiating bound dark states contribute as well by emitting dressed energy through dark-to-bright emission mechanism. All the processes modulated at sub-cycle time scale jointly form this new-type coherent XSC. This work achieves the strong-field attosecond control of the exotic atomic radiation dynamics and provides the means of simultaneous generation of separated attosecond sources, i.e., XSC and HHG, with potential advancing attosecond interferometry.
\end{abstract}

\maketitle

Coherent supercontinuum radiation provides a unique source for time-resolved spectroscopy \cite{goulielmakis2010real}, frequency comb technologies, nonlinear microscopy \cite{Zipfel2003}, and waveform synthesis \cite{wirth2011}. It is still challenging to generate coherent continuum radiation with ultrabroad bandwidth, especially in extreme ultraviolet spectral region, which is essential for isolated attosecond pulse synthesizing \cite{Huillier1993,Paul2001,Hentschel2001}.
Discovered half a century ago, supercontinuum generation has been originally referred to the dramatic spectral broadening of ultrashort light pulses propagating in media due to a catalog of nonlinear processes, such as self-phase modulation, self-focusing, stimulated Raman scattering, and self-steepening, all of which are related to the third-order nonlinearity \cite{Alfano1970,Yang1984,Corkum1986}. Nowadays, it has been extended to include very high-order and nonperturbative nonlinear optical responses for producing attosecond and zeptosecond supercontinua \cite{Hentschel2001,Paul2001,Popmintchev2012}. 

The unusual and fascinating multioctave spectral broadening challenges the understanding of light-matter interaction down to the fundamental level \cite{Elu2020,Zheltikov2019,Beetar2020}. Quantum mechanically, the photon is emitted by the transition between two states with frequency corresponding to the transition energy. When the upper states are continuum states, a continuous emission spectrum is expected, e.g., strong field ionization induced high harmonic generation (HHG) which gives rise to the emission of attosecond pulses \cite{Huillier1993,Paul2001,Hentschel2001} and thus the emergence of rapid-advancing attosecond science \cite{Krausz2009}. However, when only bound states are involved, how do we understand the generation of light continuum? Immediately, one realizes that emission (absorption as well) can occur between the photon-dressed states that is how the nonlinearity steps in \cite{Huillier1993} and how the recently advocated Floquet engineering is rooted \cite{Zhou2023}. On the other hand, when the driving pulse duration is down to few cycle or even single cycle \cite{wirth2011,Rossi2020}, which itself is synthesized from a broadband supercontinuum, the simultaneous dressing by the multiple photon modes and their interplay make the  photon-dressing picture much complicated and thus inconvenient. One has to turn to the time-domain to gain insight of high field supercontinuum generation with the sub-cycle quantum dynamics instead. 

In this Letter, we report the surprising observation of coherent XUV supercontinuum (XSC) generated from helium gases driven by intense few-cycle pulses, with bandwidth more than 20 electron volts, covering wavelength from 35~nm to 73~nm. The measured XSC spectrum spans from the below-threshold region to far above the ionization potential of helium atoms. The spectral range covers the excited and autoionization states of most atoms and molecules, which makes it an ideal multi-spectral probe of ultrafast electron dynamics in materials \cite{Wang2010,Calegari2014,Schultze2014}. The bandwidth of the XSC can be further broadened by increasing the driving laser intensity, offering a new method to produce ultrashort isolated XUV pulses. Strikingly, the XSC is emitted at a large divergence angle and consequently spatially separated from the accompanying HHG, which provides the possibility to simultaneously generate two synchronized attosecond light sources for future attosecond interferometry measurements.

\begin{figure*}[ht]
  \centering %
  \includegraphics[width=0.85\textwidth]{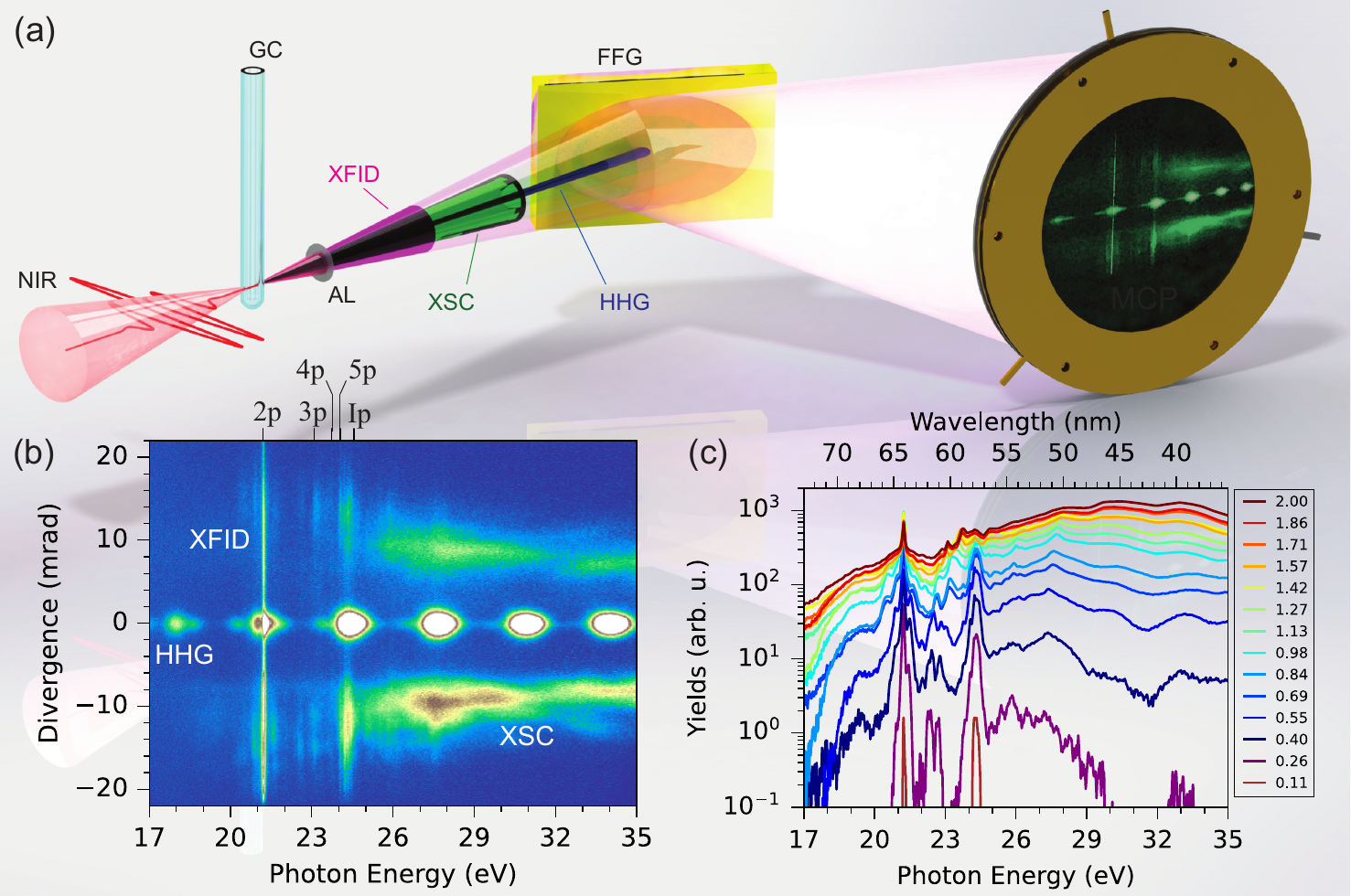}
\caption{ Illustration of the interaction scheme and spatiotemporal distribution of the emissions. (a) A NIR pulse with duration of 6\,fs was focused into the 60\,mbar helium gas cell. The emission spectra including HHG, XSC and XFID were spectrally and spatially resolved by a high-resolution XUV spectrometer, which consists of a flat-field grating and a MCP imaging detector. (b) Spatial profiles of HHG, XSC and XFID emission spectra generated with the laser intensity of 0.75~PW/cm$^2$. The discrete HHG is emitted with near zero divergence, the XFID shows line-like emission with large divergence angle ($\pm$20 mrad), and the interested XSC emission covers a broad spectrum of 17--35 eV, which corresponds to the 11st--21th harmonic orders. (c) The emergence of the XSC with increasing laser intensities. Colored lines represent the measured spectra with laser intensities of 0.1--2.0\,PW/cm$^2$. }
\label{fig1}
\end{figure*} 

The robust generation of the new type XSC calls for a new understanding of the role of bound excited states. Here we demonstrate that the atomic bound states, which have been much ignored previously in the HHG process, can be coherently populated, strongly dressed by intense few-cycle laser pulses and emitting coherent XUV supercontinuum. For helium atom, the lowest excitation energy is as high as 21.2 eV, but we found still that the excited states are significantly populated by the strong field.  After the driving pulse is turned off, the created quantum coherence by the excitation makes the atoms acting as a giant dipole oscillating in chorus giving rise to coherent emission at the resonant frequency via XUV free induction decay (XFID) which resembles the nuclear  \cite{Bloch1946}. However, during the interaction with the intense few-cycle laser pulses, the atomic resonant line emissions are found strongly modulated with the emission spectra at the transient AC Stark shifted or Floquet-dressed transition energy \cite{Delone1999}. In particular, we show that the dark states that do not participate in dipole radiation during field-free case, can be turned into bright states by Floquet dressing, which builds up the bridge for the XSC generation. 

The XSC emission and its dependence on the driving laser intensity were investigated with experimental scheme shown in Fig.~\ref{fig1} (a). A near-infrared (NIR) pulse with a duration of 6\,fs was focused into 60\,mbar helium gas cell using a concave mirror with f\,=\,500 mm. The peak laser intensity was estimated to exceed 2.1 PW/cm$^2$ and could be decreased with the combination of a broadband half plate and a broadband polarizer. The emission spectra were spectrally and spatially resolved using a high-resolution XUV spectrometer. Fig.~\ref{fig1} (b) shows the typical recorded spectra with spatial profiles from helium atoms ignited with few-cycle laser pulses at the intensity of 0.75 PW/cm$^2$. As indicated, there are three XUV emission channels: (1) the well-known HHG, which has a very small divergence angle and exhibits a frequency comb with frequency interval of twice of the central frequency of the driving pulse; (2) the XFID emissions, which exhibit sharp line-like structures with narrow widths and a wide divergence angle. The central energy of these emissions is associated with the electric dipole transitions from various $1snp$ states to the ground state, where $n (n\,\geq\,2)$ is the principal quantum number; (3) the XSC emissions demonstrated in this work. The XSC has a characteristic divergence distribution centered at ±8 mrad and spanning less than 5 mrad (full width at half maximum), which covers a broad spectral range even beyond the upper detection limit of our XUV spectrometer at high laser intensities.

To investigate how XSC is generated and evolves with the driving laser intensity, the yield of XSC was measured with various laser intensities of 0.15$\sim$2.1 ${\rm PW/cm}^2$, as shown in Fig.~\ref{fig1} (c). At the lowest intensity, only the XFID radiation is observable and can be divided into two series. One series can be assigned to $1s2p\rightarrow1s^2$ on top of the 13th harmonic; the other corresponds to the $1snp\rightarrow1s^2$ transitions with $n\approx 5 \sim 9$, although they can hardly be resolved. XFID signals from the $3p$ and $4p$ states are less visible at lower laser intensities, which has been observed and interpreted as the re-ionization of these states \cite{Yun2018}. However, this interpretation contradicts the observation of their appearance at higher laser intensities in the present experiments. With increasing laser intensity, the continuum spectra above the helium ionization potential first gradually emerge, with the cutoff position linearly proportional to the laser intensity, then the field-dressing of the bound states form the continuum emission below the threshold. At higher laser intensities, the supercontinuum spectra in the lower-energy part well expand into that below the first excitation energy. Eventually, the resonant structure of XFID diminishes, and the spectra evolve into a continuum that cover the spectral range of 17--35 eV as presented in Fig.~\ref{fig1} (c). More interestingly, additional emission peaks between $1s2p$ and the threshold appear with energy positions that linearly shift with the laser intensity.

We now investigate the role of excited states in XSC. Excited states are known to be coherently populated by an ultrashort intense laser pulse through a multiphoton-resonance transition or frustrated tunneling ionization \cite{Nubbemeyer2008}, where the tunneling electron recombines with the ion into excited neutral states. The coherence between Rydberg states and ground state forms an oscillating dipole, which induces the XFID emission \cite{Bengtsson2017,Yun2018}. Because XFID emission mainly occurs after the pulse, the emitted photon energy is exactly the field-free transition energy; hence, the emission yield directly reflects the population of the corresponding excited state. The strong field excitation and radiative decay of the excited states have been identified with HHG \cite{Beaulieu2016,Beaulieu2017,Li2015,Chini2014}.

\begin{figure}[ht]
  \centering %
  \includegraphics[width=0.5\textwidth]{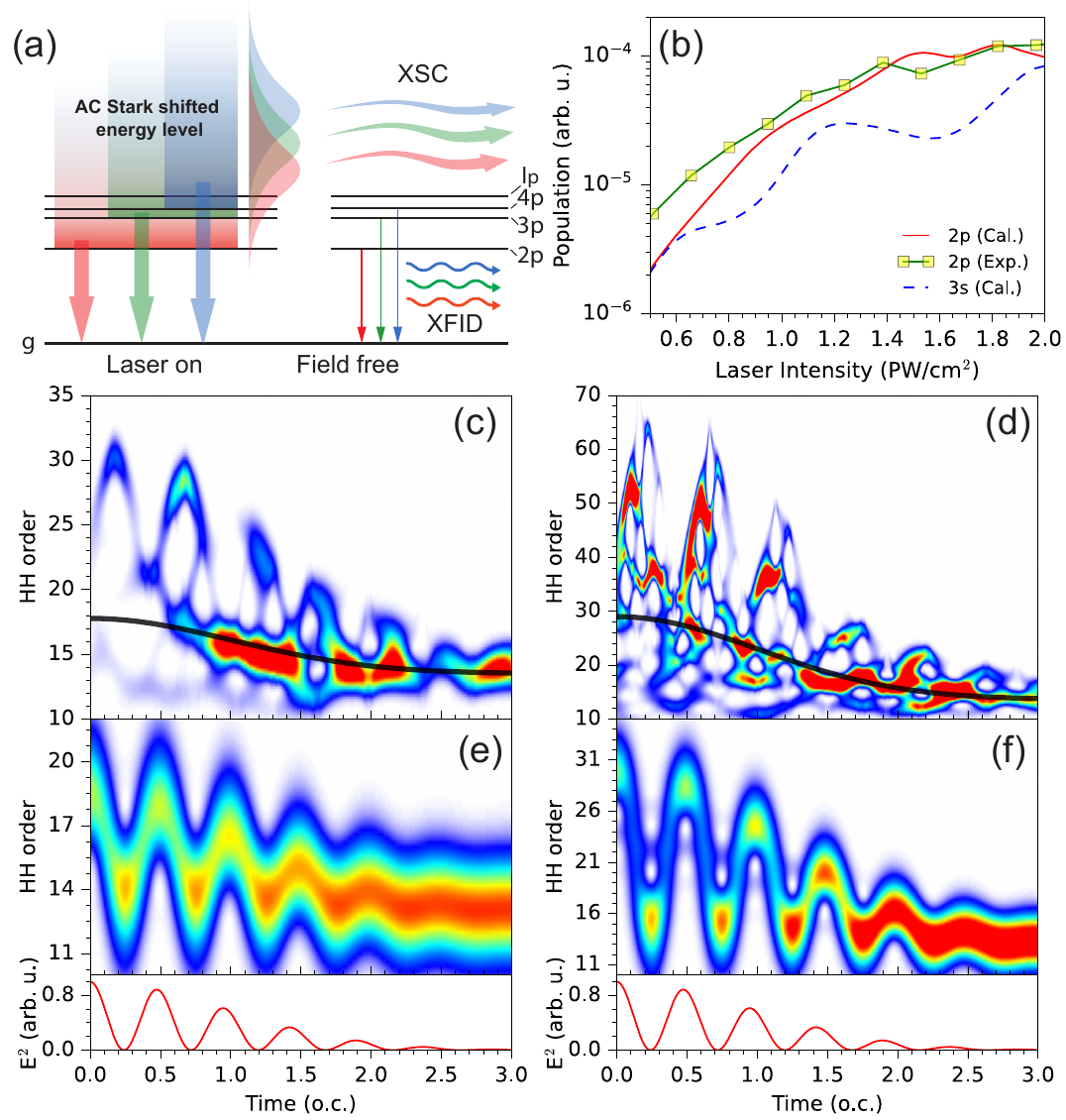}
\caption{Illustration of the XSC generation mechanism and time-frequency analysis of the coherent emission. (a) During the laser pulse, the dipole transition energies from the excited states to the ground state are continuously shifted, which induces the XSC emission. After the laser pulse is turned off, the oscillating dipole generates XFID emission at the field-free transition energy. (b) Measured $2p$ XFID emission signal and calculated $2p$ and $3s$ populations after the laser pulse as functions of the driving laser intensity. (c) and (d) Time-frequency analysis of the calculated total coherent emissions using TDSE with laser intensities of 0.1 and 0.4 PW/cm2 respectively. The black curves represent the intensity envelop of the laser pulse. (e) and (f) Coherent emissions from the oscillator model with the same laser intensities as (c) and (d). The temporal intensity profiles of the corresponding pulses are plotted in red lines at the bottom. }
\label{fig2}
\end{figure}

The energy levels of Rydberg states experience the AC Stark shift due to the strong driving laser field, so the final populations of the excited states are impacted by channel-closing effects \cite{Zimmermann2017,Liu2021} or an inter-cycle interference \cite{Xu2020,Hu2019}. As shown in Fig.~\ref{fig2} (b), the measured XFID yields from $2p$ state were modulated with the laser intensity. To understand the mechanism, we performed a simulation by solving the time-dependent Schr\"{o}dinger equation (TDSE) using a model potential with single active electron approximation, to reproduce the ionization potential and excitation energy of helium \cite{Tong1997}. Fig.~\ref{fig2} (b) plots the population of excited states $2p$ and $3s$ after the pulse as a function of the laser intensity. The population of $2p$ state is in satisfactory agreement with the experimentally measured XFID yields. Due to the selection rule for the electric dipole transition, the $ns$ states do not radiate XFID, although they are coherently populated. However, they can become bright with the help of photon-dressing as will be discussed later. The total emission spectra in Fig.~\ref{fig2} (c) and (d) are calculated by the Fourier transform of the time-dependent dipole acceleration.

The physics behind the observed XSC can be understood with the same principles as the laser-assisted continuum emission (LACE) \cite{Lei2022}, as indicated in Fig.~\ref{fig2} (a). Before the XFID emission occurs, once the $1snp$ Rydberg states are populated during the ramp-up of the laser pulse, the dipole formed by the coherent excitation of the Rydberg state and ground state remains strongly driven by the tail of the laser field. The laser-perturbed dipole radiates emission with a varying frequency, which is determined by the instantaneous Stark shift. In Fig.~\ref{fig2} (c), we performed a time-frequency analysis of the total coherent emission using the synchrosqueezing transform \cite{Li2015} at the laser intensity of 0.1 PW/cm$^2$. High-harmonic emission mainly occurs at the typical recombination instants with both long and short trajectories \cite{Salieres2001}. In contrast, the XSC continuously emits during each cycle of the laser tail with a maximum emission energy that is linearly proportional to the laser intensity profile (black curve). As the turning off of the laser pulses, the maximum emitted energy continuously decreases and eventually merges with the XFID emission. To further corroborate this mechanism, we constructed an oscillator model formed by the coherent population of the first excited state and the ground state of helium. The two states are assumed to be initially coherently populated and subsequently driven by the laser pulse. Fig.~\ref{fig2} (e) and (f) show the time-frequency property of the oscillator model with the laser intensities of 0.1 PW/cm$^2$ and 0.4 PW/cm$^2$. The spectra frequency as a function of emission time follows the temporal intensity profile of the driving pulse, which confirms that XSC is generated through the LACE process \cite{Lei2022} and indicates that the XSC emission is negatively chirped in the time domain. Unlike the molecular case where multiple vibrational states with close energies are strongly coupled by the driving laser pulse, the helium $1snp$ states are surprisingly strongly coupled to the ground state through as many as 13-photon resonances. The coherence of the excited state and ground state is indispensable for both XSC and XFID emissions.

\begin{figure}[ht]
  \centering %
  \includegraphics[width=0.45\textwidth]{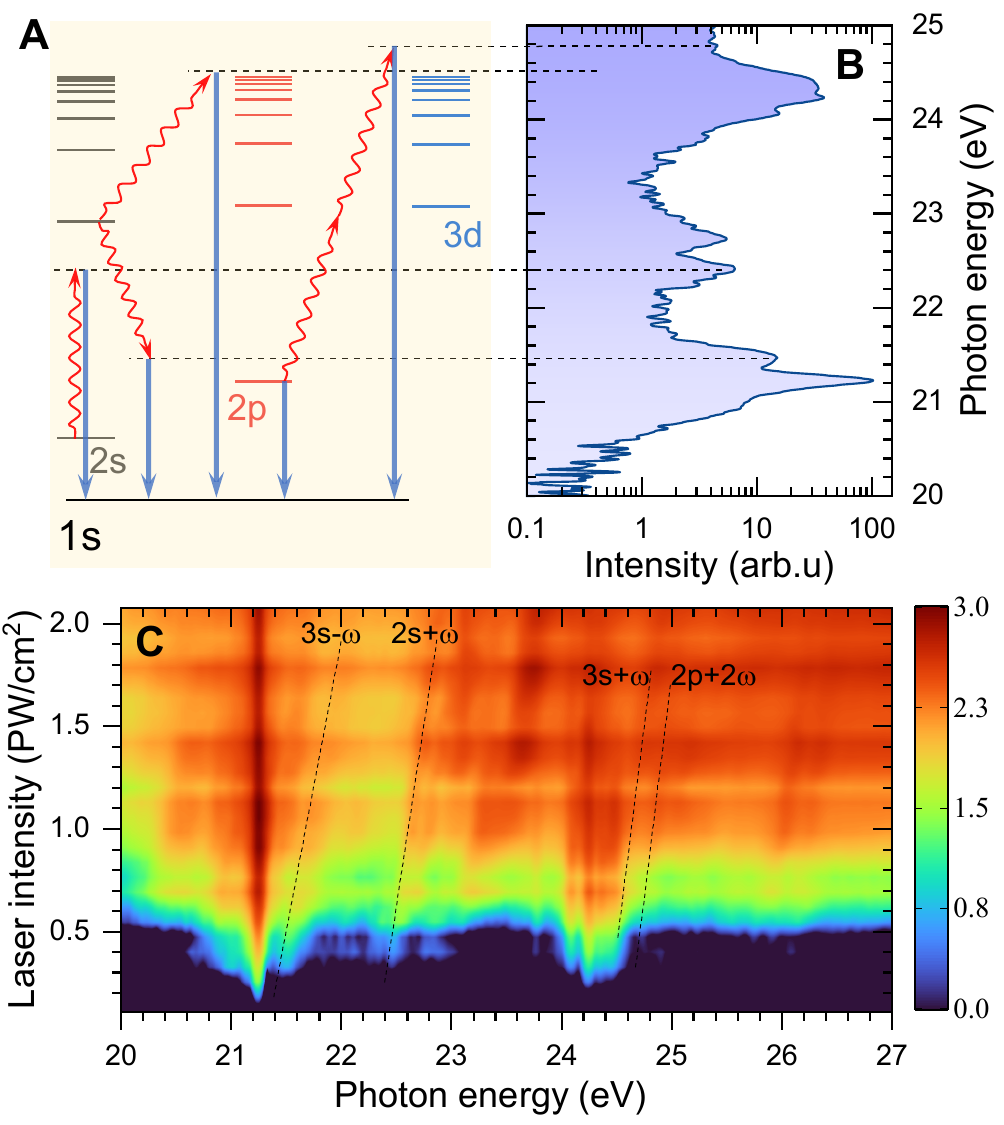}
\caption{Dark-to-bright emission. (a) Emission channels: direct transition from $2p$ and Floquet emissions $E_{2s}\,+\,\hbar\omega$, $E_{3s}\pm\hbar\omega$ and $E_{2p}\,+\,2\hbar\omega$. Here we refer the emissions from the excited states other than $p$ state are referred to as DTB emission. (b) The measured spectra can be associated with the corresponding emission channels. (c) Measured intensity-dependent emission yields. The dashed lines indicate the central energies of the DTB emissions.}
   \label{fig3}
\end{figure}

In addition to the XFID related radiative $1snp$ states, the dark states are also crucial for XSC emission. According to the selection rules, only the $1snp$ states can make direct transitions to the ground state via one-photon emission while the $1sns$ and $1snd$ states are not radiative, so they are known as dark states. However, under the strong drive of the laser pulse, they remain populated as shown in Fig.~\ref{fig2} (b) and can be dressed with additional photons to form Floquet states. Therefore, those dark states can be converted into bright states by emitting the dressed energy. For example, the $2s$ and $3s$ dark states lead to the emission of $E_{2s}\,+\,\hbar\omega$ and $E_{3s}\pm\hbar\omega$, respectively, as illustrated in Fig.~\ref{fig3} (a) and (b). We will refer to them as dark-to-bright emissions (DTB). Floquet states are reached by multiphoton excitation or frustrated ionization instead of the XUV photon-absorption that has been observed in the transient attosecond absorption spectra \cite{Chini2013,Fidler2019}. Certainly, this type of dressed emission is not limited to dark states for the $p$ states. Because these dark sates are coupled to other states, their energies experience an additional AC Stark shift, which linearly depends on the laser intensity. Therefore, the central energy of each DTB emission exhibits a linear intensity dependence, as shown in Fig.~\ref{fig3} (c), which plays essential role to bridge the energy gap among bound states in XSC. 

\begin{figure}[ht]
  \centering %
   \includegraphics[width=0.45\textwidth]{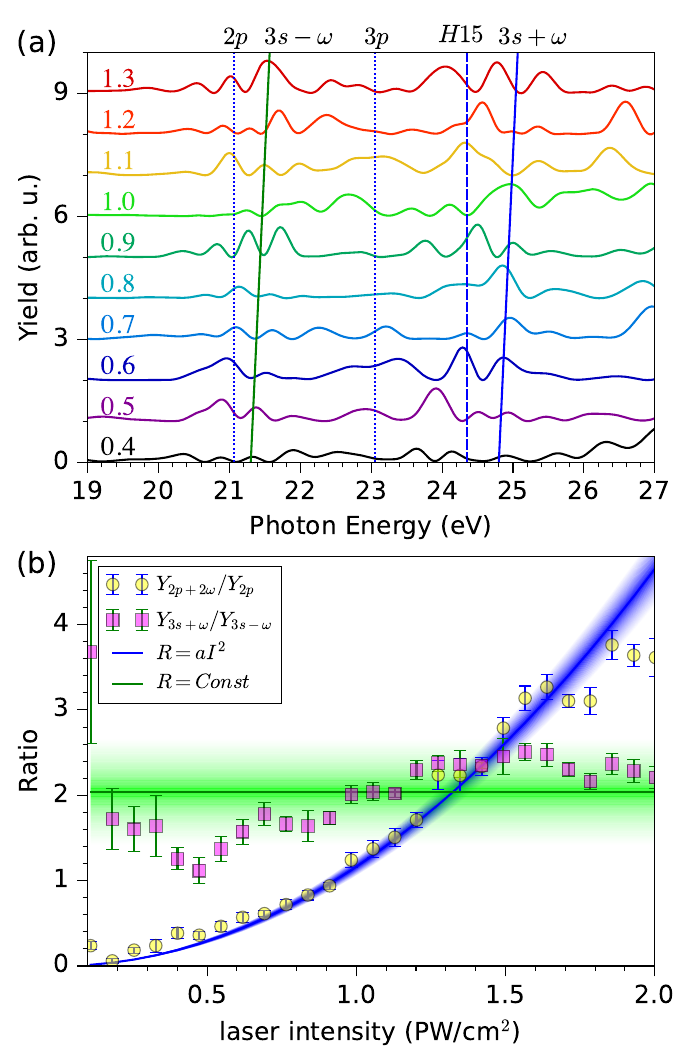}
   \caption{ Analysis of the emission spectra and intensity. (a) Simulated emission spectra for laser intensities from 0.4 to 1.3 PW/cm$^2$. The excitation energies to $2p$,  $3p$, and the 15th harmonic are indicated by vertical dashed lines. The $E_{3s}\pm\hbar\omega$ channels are identified by finding the relevant spectral peak that linearly shifts with increasing laser intensity. For the guidance, solid lines are drawn by connecting the energy positions. (b) Ratio of the measured yields: $R_{3s}=Y_{(3s+\omega)}/Y_{(3s-\omega)}$ (purple squares) and $R_{2p}=Y_{(2p+2\omega)}/Y_{2p}$ (yellow circles). The solid lines indicate that the ratio $R_{2p}$ quadratically scales with the laser intensity, whereas $R_{3s}$ is constant when the laser intensity changes.}
  \label{fig4}
\end{figure}

The TDSE simulation of the coherent emission spectra at various laser intensities are shown in Fig.~\ref{fig4} (a). The HHG signals are mixed with XSC and XFID signals. As experimentally observed, different phase behaviors of the emission as functions of the laser intensity cause their spatial separation due to phase matching. While the present simulation with only single-atom responses predicts more complex structures caused by the interference among three emission channels, which may be smeared after considering the propagation. Nonetheless, we can still identify the transition lines from $2p$ and $3p$ states to the ground state in Fig.~\ref{fig4} (a). No spectral peak appears at the $ns$ excitation energies because the direct dipole transition to the ground state is prohibited. In particular, we identify the $E_{3s}\pm\hbar\omega$ DTB channels as indicated by the solid-slash lines. Furthermore, we retrieve the measured spectral yields $Y$ along each DTB emission and take the ratio of the yields as $R_{3s}=\frac{Y_{(3s+\omega)}}{Y_{(3s-\omega)}}$ and $R_{2p}=\frac{Y_{(2p+2\omega)}}{Y_{2p}}$, as plotted in Fig.~\ref{fig4} (b). The ratio $R_{2p}$ quadratically scales with the laser intensity, whereas $R_{3s}$ keeps constant when the laser intensity changes. Thus, the simulation and measurement confirm the existence of DTB emission.

During the interaction between few-cycle strong laser pulses and atoms, HHG, XFID and XSC are generated simultaneously. However, the three types of emissions can be spatially separated and temporally distinguished as illustrated in Fig.~\ref{fig1} (a): HHG occurs within the driving pulse, followed by the emission of XSC at the tail of the pulse and emission of XFID after the pulse. Besides, they are also spatially separated. The high harmonics are propagated with the driving pulse under phase-matching. The coherence between Rydberg states and ground state forms the oscillating dipole with a large phase acquired from the laser-driven electron dynamics, which leads to a large divergence emission of XFID after the pulse. The XSC is emitted through the LACE mechanism, i.e., the excited helium atom is coherently polarized during the tail of the laser pulse, which causes a cone-like emission due to the off-axis phase-matching \cite{You2012}. 

In summary, we uncovered the interplay of the three-emission mechanism of helium pumped by intense ultrashort laser pulses. We observed the coherent supercontinuum emission in the XUV region and identified their mechanism, spatial behaviors and temporal behaviors. It is demonstrated that the quantum coherence created and driven by the intense ultrashort pulses results in an emission at the AC Stark shifted or Floquet-dressed transition energy, which differs from the conventional nonlinear optics and the well-studied HHG. Our findings shed a new light on the role of non-radiative dark states which participate in XSC generation. The spatio-temporal specificity of the supercontinuum enables simultaneous generation of XSC attosecond pulse and accompanies HHG attosecond pulses, which can be potentially applied for attosecond interferometry.

\begin{acknowledgments}
This work was supported by National Key Research and Development Program of China (Grant No. 2019YFA0307703), and the National Nature Science Foundation of China (Grant Nos. 12234020, 12274461, 11904400).

\end{acknowledgments}

\bibliography{XLACE}

\end{document}